# A transferable MARTINI model of polyethylene oxide

Fabian Grunewald[1], Giulia Rossi[2], Alex H. de Vries[1], Siewert J. Marrink[1], Luca Monticelli[3],*

[1] Groningen Biomolecular Sciences and Biotechnology Institute and Zernike Institute for Advanced Materials, University of Groningen, Groningen, The Netherlands

[2] Physics Department, University of Genoa, via Dodecaneso 33, 16146 Genoa, Italy

[3] Molecular Microbiology and Structural Biochemistry, UMR 5086, CNRS and University of Lyon, France

*Supporting Information Placeholder*

**ABSTRACT:** Motivated by the deficiencies of the previous MARTINI PEO models in apolar environments, we present a new PEO model based on (a) a set of 8 free energies of transfer of dimethoxyethane (PEO dimer) from water to solvents of varying polarity; (b) the radius of gyration of PEO-477 in water at high dilution; and (c) matching angle and dihedral distributions from atomistic simulations. We demonstrate that our model behaves well in five different areas of application: (1) it produces accurate densities and phase behavior or small PEO oligomers and water mixtures; (2) it yields chain dimensions in good agreement with experiment in three different solvents (water, diglyme and benzene) over broad range of molecular weights (~1.2 kg/mol to 21 kg/mol); (3) lipid bilayers with PEGylated lipids based on our new model reproduce qualitatively the structural features in the brush and mushroom regime; (4) the model is able to reproduce the phase behavior of several PEO-based non-ionic surfactants in water; (5) it can be combined with the existing MARTINI PS to model PS-PEO block-copolymers. Overall, the new PEO model outperforms previous models and features a high degree of transferability.

## 1 - Introduction

Polyethylene glycol (PEG), also known as polyethylene oxide (PEO), is one of the few polymers with an exceptionally wide scope of applications ranging from bio-medical applications, over cosmetics and food additives to the active material in polymer batteries. Many applications of PEO involve multiple chemical components and supra-molecular assemblies in non-crystalline phases, for which structural information is typically only available at low resolution (if at all). Molecular dynamics (MD) simulations are a powerful method to gain an insight into the structure and dynamics of liquids and soft matter, including biological macromolecules and polymers.

Depending on the time and length scales relevant for the specific system at hand, either atomistic or coarse-grained (CG) molecular dynamics simulations can be used to characterize and even predict the properties of materials. One of the most commonly used CG models for biomolecular simulations is the MARTINI model. MARTINI is based on a building block approach, i.e., each building block represents a chemical moiety and is parameterized separately; larger molecules are obtained by stitching together multiple building blocks. MARTINI represents a group of about 4 heavy atoms as one particle (bead). Each bead has a specific type, which is defined by a set of Lennard-Jones potentials for the interaction with the other beads in the force field. Electrostatic interactions are calculated for particles holding a full charge. The choice of the bead type for a given group of atoms is based on matching the free energies of transfer of the chemical moiety with experimental data.[1]

MARTINI has been used successfully to model a range of polymers[2–6], and several MARTINI models have been published also for PEO[7–11]. Yet, it was realized already in the first published parametrization that none of the standard MARTINI beads is appropriate for modeling a PEO type repeat unit because some structural and thermodynamic properties could not be matched accurately enough.[12] As similar observations were made for other models, new beads with custom made interactions were introduced on several occasions.[7,8,11] Usually the authors intended to approach a specific problem, so the new PEO beads were optimized to reproduce some specific property of the system of interest. Some of these initial PEO parameterizations have since been refined in multiple steps to extend their scope. From here onwards, we shall define a model as a parametrization of PEO, including a new bead type, for which an interaction matrix with all other MARTINI beads is provided.

The first model of PEO was put forward by Lee *et al.*[12], shortly followed by the model of Rossi and coworkers.[8] The Lee model started from one of the standard MARTINI beads (SNa), and the authors found that the radius of gyration and end-to-end distance could only be reproduced by using a SNda bead type. However, they anticipated that the rest of the interactions were more appropriately represented by the SNa type. Thus it was proposed to take the self-interaction and the interaction with water from the SNda type, and interactions with other particles from the SNa type, effectively creating a new bead.[12] In a later study based on PEGylated lipids, the SNa was then changed to SN0 to reduce the excessive adsorption of PEO tails onto lipid head groups.[7] The last refinement on the model was completed in 2013 with the introduction of new bonded parameters, which reduced the instability the model suffered from due to its dihedral potentials in the backbone of the polymer.[9]

In contrast, the Rossi model was aimed at reproducing the experimental free energy of transfer of dimethoxyethane (PEO dimer) from water to octanol. Since none of the standard beads was able to yield a sufficiently accurate free energy of transfer, the authors decided to create a new bead. The self-interaction was subsequently determined from the long-range structural properties. Since the new bead was an intermediate to the standard Nda and P1, the rest of the interaction table was provided based on similarity. As the Rossi model had no dihedral potential, its numerical stability was superior to the Lee model.[8]

While both models proved successful in their special cases and have been reused in similar environments, transferability to different chemical environments remained problematic. In particular, the interaction with hydrophobic phases was much too unfavorable for both models. First Huston and Larson[13] pointed out that the behavior of PEO at hydrophobic interfaces is incorrect for both models. Later Carbone et al. noticed that the Rossi model displays too collapsed conformations in hydrophobic solvents.[6] It was already realized earlier that the water-hexadecane free energy of transfer was far off for both models compared to an estimate based on experimental data.[14] Later, free energies of transfer obtained from atomistic simulations confirmed that both PEO models were too hydrophilic by a large amount.[11] The excessive hydrophilic character made both models problematic to use in non-polar environments, which are relevant in the field of materials science – for example, lithium ion batteries, where polystyrene (PS)-PEO copolymers are self-assembled in apolar solvents.[15–17]

Here we present a new model for PEO, characterized by a high transferability between different environments, especially extending the domain of usage to non-polar solutions. The new model is based on reproducing free energies of transfer of dimethoxyethane, obtained either from experiment or from atomistic simulations, and it is applicable over a wide range of molecular weights, spanning three orders of magnitude. We show that the new model reproduces essential features of previous models and improves on their results; models of PEGylated lipids show reasonable performance, and excellent agreement with experimental data is obtained for the phase behavior of nonionic surfactants. In addition, we show that the new model can be combined with the current MARTINI polystyrene (PS) model to give one of the technologically most relevant block-copolymers, PS-PEO.

## 2 – Methods

As in previously published polymer models[2,4,8,18] the parameterization of the new PEO model is based on (A) free energies of transfer of dimethoxyethane between a range of solvents, and (B) long range structural properties of isolated polymer chains – namely the radius of gyration of a long polymer chain (477 residues) in water, calculated at high dilution. Target values for the free energies of transfer and the radii of gyration are taken from experiments whenever available; when unavailable, target values are calculated from atomistic simulations. Below we describe first the setup and parameters for free energy calculations and simulations of individual chains in solution; then we report the methods used for validation of the new model (radius of gyration in different solvents, the phase behavior of PEO oligomers, PEGylated lipids, non-ionic surfactants, and PS-PEO micelles).

### 2.1 - Free Energy Calculations

Free energies of transfer of dimethoxyethane were calculated at the atomistic and coarse-grained level as differences between free energies of solvation in different solvents. Solvation free energies were computed by alchemical free energy transformations as implemented in the GROMACS package. The free energy of the transformation was estimated using the Multi-state-Bennetts-Acceptance-Ratio (MBAR) method,[19] obtained using a python tool available on github (https://github.com/davidlmobley/alchemical-analysis). For each calculation, the convergence and quality of the calculations were checked following the guidelines suggested by Klimovich, Shirts and Mobley.[20] The error reported with the calculations is the statistical error estimate. For both sets of simulations, the intra-molecular interactions were not switched off.

#### 2.1.1 – Atomistic calculations

All atomistic simulations were run using the GROMOS 2016H66 force field. This force field has been validated against bulk properties of many solvents[21] and has ether oxygen parameters, which have been shown to reproduce correct solvation free energies for dimethoxyethane[22], as well as a correct phase behavior for non-ionic surfactants[23], of interest for the present work.

Lennard-Jones (LJ) and Coulomb interactions were cut-off at 1.4 nm, which is the standard GROMOS cut-off. Long-range Coulomb interactions were treated by the reaction field approach, with the relative dielectric constant set to the value of the bulk solvent (also following the GROMOS standard treatment). All bond lengths were constrained, as in the original work.[21] Our simulation conditions differ from the standard GROMOS conditions in two respects: (1) we ran all simulations with the GROMACS software, which implements a highly efficient Verlet cut-off scheme (Verlet buffer tolerance of $10^{-6}$ kJ/mol/ns per particle) for the non-bonded interactions instead of the standard GROMOS twin-range cut-off (used in the GROMOS parameterization and unavailable in GROMACS); (2) LJ and Coulomb modifiers were used to shift the potential to zero at the cut–off, to avoid discontinuities in the potential and improve energy conservation. To verify that our settings reproduce GROMOS results and yield acceptable solvent properties, we calculated the density and heat of vaporization of each bulk solvent (see supporting information S1). Our results compare very favorably with the values from the original publication and experiment, with deviations in the heat of vaporization below 1.9 kJ/mol in both cases. The force field files, including run parameters and starting structures, are available online (github and MOBI website, http://mmsb.cnrs.fr/en/team/mobi).

Following the recommendations by Klimovic et al.[20], two different sets of simulation parameters were employed for the simulations of bulk solvent with and without partial charges. In the GROMOS model, propanethiol, butanol, acetone and propanol have partial charges. In this case, the lambda vector for switching off the interactions was split into its Coulomb and Lennard-Jones components; first, we switched off the Coulomb interactions between solvent and solute, then the LJ interactions. For the simulations in cyclohexane, octane and benzene, which have no partial charges, only one lambda vector was used. In order to improve convergence, in both sets of simulations soft-core potentials were employed, using the parameters detailed by Shirts and coworkers.[24] Each window was run for either 16 ns or 20 ns, and a variable amount of equilibration time was discarded based on convergence analysis (following Klimovich and coworkers[20]). The derivative of the potential energy with respect to lambda was computed every 50 steps.

All simulations were performed using the stochastic dynamics (SD) integrator implemented in GROMACS (version 2016.4), with a time step of 2 fs. Production runs were performed in the NpT ensemble at 298.15 K (with inverse friction constant of 2 ps), using the Parrinello-Rahman barostat to fix the pressure at 1 bar (time constant of 2 ps).

### 2.1.2 - Coarse-grained calculations

For the coarse-grained (CG) simulations, the MARTINI force field version 2.2 was used as available online (http://mmsb.cnrs.fr/en/team/mobi). The run settings were the same as suggested by de Jong et al.[25] (cut-off for non-bonded interactions: 1.1 nm; Verlet neighborlist scheme) with the exception of the verlet-buffer-tolerance, which was decreased from the GROMACS default value to $10^{-6}$ kJ/mol/ns per particle.

Since none of the MARTINI models in the system of interest have partial charges, only one lambda vector of 15 non-uniformly spaced points was used to switch off the LJ component of the potential energy. All 15 windows were run for 16 ns and the derivative with respect to lambda was computed every 10 steps.

All CG simulations were carried out with the GROMACS software (version 2016.4), using the stochastic dynamics integrator (with inverse friction constant 1.0 ps) and a time step of 20 fs. Production runs were carried out in the NpT ensemble at 298.15 K and 1 bar (time constant 4.0 ps and compressibility 4.5 $10^{-5}$ bar$^{-1}$).

## 2.2 - CG simulations of PEO systems

We used our newly developed PEO model for all simulations of PEO systems. All topology files and starting structures were generated using the python tool Polyply, which can generate starting structures and topology files (compatible with the GROMACS software) for both atomistic and coarse-grained polymer chains. The tool will be described in detail in a separate publication (manuscript in preparation), and a preliminary version including instructions can be found on GitHub (https://github.com/fgrunewald/Martini_PolyPly).

### 2.2.1 - Single Chain in Solution

The radius of gyration and end-to-end distance of PEO in three different solvents (water, benzene and diglyme) was obtained by simulating a single chain in a box of solvent. For the first two solvents, the temperature was fixed at 298.15 K using the velocity rescale thermostat introduced by Bussi and coworkers.[26] In contrast the simulation in diglyme was performed at 323.5 K, which is the theta temperature of this solvent.[27] For all simulations, the pressure was fixed at 1 bar using the Parrinello-Rahman barostat (time constant of 10 ps). For each solvent, 5 different molecular weights were considered, from about 1.2 kg/mol to 11 kg/mol; in the case of water, one additional simulation with a molecular weight of 21 kg/mol (corresponding to 477 monomers) was performed. All simulations in water and diglyme were run for at least 30 µs, while the simulations in benzene were run for at least 20 µs. For all systems of PEO the standard GROMACS MD integrator with a time step of 20 fs was used.

To avoid artifacts from periodic boundary conditions and interactions between periodic images, all simulations were conducted in the dilute regime, at concentrations below the approximate overlap concentration. The overlap concentration is given by[28]:

$$\varphi^* \approx \frac{N \times b^3}{<R>^3} \approx \frac{1}{N^v}$$

In this case N is the number of repeat units, with length b, of an equivalently jointed chain. $<R>$ is the end-to-end distance. The exponent v is approximately 0.5 for theta solvents and 3/5 for good solvents. N can be approximated from the characteristic ratio, as explained in Supporting Information.

### 2.2.2 - Solutions of PEO oligomers

Solutions of PEO oligomers in water were simulated at 298.15 K and 1 bar pressure, using the same run parameters as for the simulation of single chains in solution. For the four oligomers dimethoxyethane (DXE), diglyme (DEG), triglyme (TIG) and tetraglyme (TRG) the density was computed from simulations of 200 ns (previously equilibrated for 12 ns with Berendsen pressure coupling).

### 2.2.3 – PEGylated lipids

Two bilayer systems containing mainly DPPC and smaller amounts of DOPE as well as PEGylated DOPE (PEL) (see table 1) were simulated at 283 K. Each simulation was prepared by first generating the bilayer using the python tool insane.py[29], and contained 1 DOPE lipid in each leaflet and the appropriate number of DPPC lipids to reach the desired concentration. Subsequently, PolyPly was used to grow a 45-repeat unit PEO chain onto one of the DOPE lipids. PEO chains were terminated by one SP2 bead (to represent the terminal hydroxyl group). Details on mapping and bonded interactions are provided in the next section. Afterwards the system was stacked in the xy-plane to obtain the final bilayer. Note that only one leaflet contained the PEGylated lipids. This choice was made to ensure the PEO tail does not interact with its periodic image. The equilibrium box dimensions were 21.72 nm by 21.72 nm by 37.60 nm for system A and 18.57 nm by 18.57 nm and 50.01 nm for system B, respectively. Note the higher amount of water (normal W and anti-freeze WF beads) in system B due to the expectation of a more stretched chain. The simulations were run for about 4 µs. The run parameters were the same as those used for the radius of gyration simulation, with the exception of the pressure coupling scheme (semi-isotropic instead of isotropic).

**Table 1. Composition of bilayers with PEGylated lipids**

| Molecule | # in bilayer A | # in bilayer B |
|----------|----------------|----------------|
| DPPC     | 2016           | 864            |
| DOPE     | 16             | 144            |

| PEL/Na+ | 16 | 144 |
| W | 125,984 | 127,872 |
| WF | 1600 | 1440 |

### 2.2.4 – Nonionic Surfactants

The self-assembly of 3 types of nonionic surfactants (C12E6, C12E4, C12E2) mixed with water at three different concentrations (50w%, 53w%, 71.15 w%,) were simulated. The run parameters were the same as those used for measuring the radius of gyration. However, the pressure coupling in this case was done using a semi-isotropic Berendsen barostat with both the z and xy component of the pressure fixed to 1 bar using a coupling time of 2 ps and a compressibility of 4.5 bar$^{-1}$. The surfactants and water molecules were coupled separately to the thermostat using a coupling time constant of 4 ps and a reference temperature of 298.15 K. Each system was run 6 times, each time with a different 6 digit long random-seed for generating random velocities from the same initial structure. Each simulation was run for 5 µs, to ensure the observed structures were stable in time.

### 2.2.5 – PS-PEO Micelles

A system containing 370 and 740 oligomers of the block-copolymer PS-PEO with lengths of 10 and 23 repeat units, respectively, was simulated at 298.15 K and 1 bar pressure, using the same simulation parameters as for the radius of gyration simulations except for the pressure coupling (Berendsen instead of Parrinello-Rahman barostat). The initial structure of PS-PEO was generated using PolyPly. A single chain was equilibrated in water, then the systems of interest were generated by inserting 370 or 740 copies of the polymer chain at random positions with random rotation into a box, and solvating with water (182,942 and 388,630 MARTINI water particles respectively). The dimensions of the final box sizes were 28.4x28.40x28.4 and 36.4x36.4x36.4 nm. The simulations were run for 3.4 µs.

The dimension and aggregation number were determined using a homemade python script, which utilizes the scikit-learn[30,31] library implementation of DBSCAN[32] to cluster beads of PS-b-PEO into aggregates based on the number density. Reading and processing of topology and trajectory information is done with MDAnalysis.[33,34] Details on the procedure for computing the radius of gyration and aggregation number are outlined in the supporting information. The script is available online free of charge (https://github.com/fgrunewald/tools_for_MD_analysis).

### 2.2.6 Assessment of convergence and error estimation

Assessment of convergence and error estimation is crucial when determining any property of polymer chains (e.g., radius of gyration, end-to-end distance, etc.). To ensure reproducibility, a three-step protocol was used to assess convergence: (1) average properties were plotted as a function of the fraction of total simulation time; (2) the same was done for the autocorrelation time (estimated with a procedure proposed by Chodera and coworkers[19,35]); (3) the autocorrelation time was also estimated using the block averaging approach described by Hess.[36] The error was estimated from the uncorrelated data set after subsampling the original data using the pymbar suit (https://www.github.com/choderalab/pymbar). All analyses were carried out with a python tool provided online on GitHub (https://github.com/fgrunewald/tools_for_MD_analysis); details on its usage and on the analysis of convergence and error estimation are reported in Supporting Information (S.2).

## 4.0 - Results and Discussion

### 4.1 - Parametrization of PEO and associated compounds

In this section, we present the parameters for the new PEO model as well as for related compounds: nonionic surfactants, polystyrene (PS)-PEO block copolymers, and PEGylated lipids.

#### 4.1.1 - Mapping Schemes

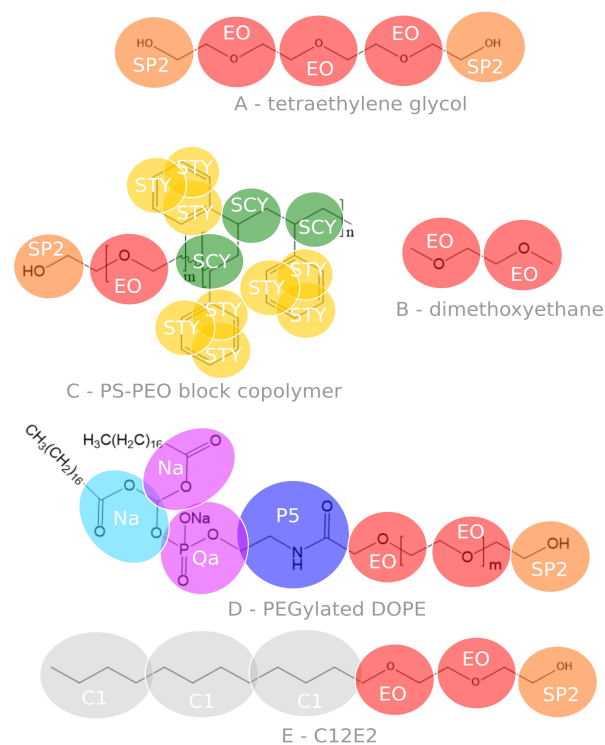

Figure 1. Representation of PEO and PEO-containing compounds each different bead type is indicated by a different color with the type displayed within the bead.

The mapping procedure consists in selecting groups of atoms and representing them by one interaction center (bead). In MARTINI, the number of atoms per bead usually varies between three and five and there are no strict mapping rules. Hence there can be several equally valid mappings for the same molecule. After the mapping scheme has been defined, the interactions between the different beads are chosen from an interaction matrix based on reproducing the free energy of transfer of the individual beads (or related compounds). The previous models of PEO have essentially used the same mapping scheme, with differences in the way end-groups were treated. In this mapping scheme, a PEO repeat unit consists of the sequence -[CH$_2$ - O - CH$_2$]- as opposed to the definition of a repeat unit in polymer chemistry textbooks, which usually is -[O - CH$_2$ - CH$_2$]-.[37,28] There are two distinct advantages of using the first representation: first, there is a more chemically intuitive connection between the small-molecules dimethylether (DME) and dimethoxyethane (DXE), representative of the monomer and dimer of the repeat unit. Second, the same mapping scheme has been used before, therefore, one can hope to retain much of the previous parametrization in terms of bonded interactions. However, disadvantage arise with respect to (1) the way the length of the polymer chain is defined, and (2) the way end groups are treated. The first mapping scheme is not fully commensurate with the underlying atomistic structure.

For instance, compound A in Figure 1 (tetraethylene glycol) is a PEO tetramer; with the first mapping scheme, we can define three repeat units; this, however, suppresses two terminal CH$_2$OH groups. Lee et al.[12] suggested to neglect such detail, and simply add an additional bead of the same type, so that an n-mer of PEO consists of n beads of the same type. While this choice is intuitive and a good enough approximation for long chains, it reduces the polarity of shorter chains (hydroxyl groups are significantly more polar than ether groups).

To take into account the higher polarity of OH-terminated chains, it is possible to add one SP2 bead at the chain end. While mapping two heavy atoms into one bead is unusual for MARTINI, it has previously been shown that including a more polar end-group is crucial for obtaining the correct phase behavior of nonionic surfactants.[8] Furthermore, this choice improves the properties of small oligomers. Thus, we will represent the tetramer of PEO by five beads: three of one type, which we call EO, and two of the polar SP2 type. In general, for OH terminated chains, we will use n-1 PEO beads and two SP2 beads. In contrast, for methyl-terminated chains (such as DXE in Figure 2B), only EO beads are used.

For cases where another end group or possibly another polymer is attached to one end of the chain, as is the case for nonionic surfactants (Figure 1E), or a PS-PEO block-copolymer (Figure 1C), the PEO part of length n will contain n beads of type EO plus one SP2 end-group and the rest of the molecule. For example, the surfactant C12E2, shown in figure 1E, contains 12 carbon atoms and two PEO repeat units, which are OH terminated. Thus, we will represent this molecule by three normal beads of type C1, two EO beads and the SP2 end group. The same reasoning can be applied to PS-PEO block copolymers (Figure 1C). We notice that, if the linking unit contains more polar atoms, a different approach might be needed.

**4.1.2 - Non-bonded interactions for the PEO type bead**

The MARTINI force field uses Lennard-Jones potentials to model non-bonded interactions. In the current version of MARTINI (v2.2), the Lennard-Jones $\varepsilon$ parameter (related to the minimum of the potential) can assume 10 different values, while $\sigma$ (related to the size of the particle) can only take two values: 0.47 nm (used for standard beads, representing about 4 heavy atoms on a linear chain); and 0.43 nm (used for ring and small beads, representing less than 4 heavy atoms). The interaction between standard and small beads has a $\sigma$ value of 0.47 nm. Since EO beads represent three non-hydrogen atoms, they should be considered as small beads ($\sigma$=0.43); this leaves the values of $\varepsilon$ as the parameter to be adjusted to reproduce the free energies of transfer.

As detailed in the introduction, the repeat unit of PEO is poorly represented by any standard MARTINI particle type. Moreover, PEO models using non-standard particle types (e.g., the Lee model and the Rossi model) are too hydrophilic.[11,13,14] In such models, the free energy of hydration of dimethoxyethane (DXE) is equal to or lower than the experimental value (-20.2 kJ/mol[38]). In contrast, the free energy of hydration for all standard MARTINI beads is higher (more positive) than observed in experiment. For these models, as a consequence, to match partitioning of DXE between water and octanol, the interactions of the non-standard PEO beads with hydrophobic particles needed to be less attractive. This caused a shift of the interaction matrix with respect to the other MARTINI beads, artificially enhancing the hydrophilicity.

**Table 2. Free energies of transfer of dimethoxyethane (DXE) from different solvents to water. The values reported for the Lee model and the Rossi model were calculated in the present work, using the published models. Reference values are taken from experiments or from atomistic calculations. Shaded solvents were used as targets, the others as validation.**

| Solvent | Bead Type | Reference | Lee et al. | Rossi et al. | this work |
|---|---|---|---|---|---|
| Octane | C1 | -7.3 ± 0.3* | -13.96 ± 0.06 | -18.86 ± 0.06 | -7.75 ± 0.06 |
| Cyclohexane | SC1 | -6.8 ± 0.3* | -24.87 ± 0.07 | -29.64 ± 0.07 | -8.66 ± 0.07 |
| Octanol | P1-C1 | -1.2† | -0.15 ± 0.07 | 0.18 ± 0.07 | -1.11 ± 0.06 |
| Benzene | SC5 | 0.2 ± 0.3* | -9.72 ± 0.08 | -14.94 ± 0.08 | 0.24 ± 0.08 |
| Propanethiol | C5 | 2.2 ± 0.3* | 1.79 ± 0.06 | -3.28 ± 0.06 | 2.03 ± 0.06 |
| Acetone | Na | 1.2 ± 0.3* | -1.53 ± 0.06 | 12.09 ± 0.07 | 1.04 ± 0.06 |
| Butanol | Nda | -2.2 ± 0.3* | -5.39 ± 0.07 | 8.91 ± 0.07 | -3.05 ± 0.07 |
| Propanol | P1 | -1.5 ± 0.3* | -3.92 ± 0.07 | 8.70 ± 0.07 | -3.15 ± 0.07 |

* from atomistic simulations and † from experiment

The new model developed here is also based on matching the free energies of transfer for DXE between different solvents. Only one experimental value for water-solvent partitioning is reported in the literature – the one for water-octanol. Therefore, in this work, we used free energies of transfer calculated from atomistic simulations as a reference for our parameterization. The solvents we chose span the entire MARTINI interaction matrix, from very hydrophilic to very hydrophobic. In this way, one can make sure not to fall victim to the same trap as the previous models did. The reference atomistic force field used was a special variant of GROMOS named 2016H66, which has been optimized with respect to ether properties and includes a sufficiently large number of well parametrized solvents.[21] The individual solvation free energies and starting setups for the calculations are available as Supporting Information

The free energies of transfer for DXE are reported in table 2. Here we make a few remarks. First, DXE prefers water over hydrocarbons, in agreement with chemical intuition and experimental observations.[reference] However, benzene is a special case: the free energy of transfer from water is about 0, meaning that DXE does not have a preference between benzene and water. While this seems quite counterintuitive at first, it is well documented that benzene is a good solvent for PEO.[39] DXE has a preference for acetone and propanethiol over water, while it prefers water over short chain alcohols.

Comparison of the reference free energies with the results obtained for the Lee and Rossi models shows that, in most cases, partitioning was not reproduced very well, with the exception of water-octanol. In this case, the underestimation of the interaction with the alkane chain is compensated by the overestimation of the interaction with the more hydrophilic components.

Since free energies of transfer are just differences between free energies of solvation, an absolute reference is also needed in order to define all interactions. One possibility is to choose the free energy of hydration as an absolute reference, as suggested by Carbone and coworkers.[11] However, in MARTINI, free energies of hydration are generally higher (less negative) than the experimental ones. Matching experimental values would be very simple, and would imply setting stronger interactions (higher values of the Lennard-Jones ε) all across the MARTINI table. The consequence of such choice would be that liquids with strong inter-molecular interactions (e.g., all polar ones) would become solid at room temperature. Considering all this, it is clear that matching free energies of hydration should be avoided (a) to maintain consistency with the rest of the MARTINI model without a complete reparameterization of the force field, and also (b) to avoid freezing of all polar liquids at room temperature. We set a value of 3.5 kJ/mol for the interaction of PEO with water (the same value as for N0), resulting in a hydration free energy of -14.73 ± 0.05 kJ/mol, which is 5.5 kJ/mol higher than the experimental value. With this choice for the water interaction, the other ε-values of the new EO bead were obtained by iteratively computing the free energies of transfer between water and selected solvents, and adjusting the epsilon value to yield the best possible agreement with the reference values. The interactions with C1, SC1, C5, SC5, Na and P1 were parametrized by matching the free energies of transfer from water to octane, cyclohexane, propanethiol, acetone and ethanol, respectively. The rest of the interaction table (Table 3) was filled in by interpolation, using the same interaction for similar bead types. This approach was validated by verifying the free energies of transfer from water to octanol and butanol; these were not used as targets in the parameterization, and yet the agreement with the reference values is very good.

The s-versions of each bead are obtained by scaling the ε-value of the normal bead by 0.75, except for the case of benzene and cyclohexane, for which a scaling factor of about 0.9 was required to obtain good partitioning free energies. Overall the free energies of transfer improve greatly from our model to the previous models, especially in the case of benzene and octane.

The self-interaction of the PEO bead was fit to reproduce the experimental radius of gyration of a single chain in water. To obtain a most reliable result, we chose a chain length of 477 repeat units (corresponding to a molecular weight of about 21 kg/mol), because scattering data is available for this chain length. Moreover, at such long chain length, the effect of the end-groups is negligible. Since the radius of gyration also depends on the bonded interactions, the self-interaction was optimized by trial-and-error in several cycles alongside the bonded interactions. In the final iteration, the value of the self-interaction was set to 3.4 kJ/mol. This yields a radius of gyration of 6.6 ± 0.2 nm, in excellent agreement with the experimentally determined value of 6.5 nm.[40]

**Table 3. Interaction matrix of the new PEO bead.**

| Bead | ε (kJ×mol$^{-1}$) | Bead | ε (kJ×mol$^{-1}$) |
| --- | --- | --- | --- |
| Qda* | 3.5 | Nda* | 3.1 |
| Qd* | 3.5 | Nd* | 3.1 |
| Qa* | 3.5 | Na* | 3.1 |
| Q0* | 3.5 | N0* | 3.1 |
| P5* | 3.5 | C5‡ | 2.95 |
| P4* | 3.5 | C4* | 2.95 |
| EO† | 3.4 | C3* | 2.95 |
| P3* | 3.1 | C2* | 2.70 |
| P2* | 3.1 | C1‡ | 2.53 |
| P1* | 3.1 | | |

\* for small beads ε = ε × 0.75

† only as small bead with ε = ε × 0.75

‡ for small beads ε = ε × 0.90

### 4.1.3 - Bonded Interactions for PEO

For the bond between two PEO beads, a simple harmonic potential was used with a reference length set to 0.322 nm and a force-constant of 7000 kJ/mol, following the original values from the Rossi model.[8] As described by Rossi *et al*. and verified here, this bond length results in better properties for the nonionic surfactants compared to the bond length of 0.33 nm used in the Lee model.

The angle and torsion potentials were optimized to reproduce the atomistic distributions of GROMOS 53A6oxy PEO in water. The target distributions were taken from the paper of Rossi and coworkers.[8] Moreover, we aimed at having high numerical stability, even for long chains, with an integration time-step of 20 fs. It has been noticed previously that MARTINI models employing a dihedral potential along the backbone may have stability problems when one of the angles approaches the value of 180 degrees. To solve this stability issue, we used the "restricted bending" potential developed by Bulacu and coworkers.[9] Figure 2 shows the distributions of the angle and dihedral for the atomistic and the CG representation. The CG distribution for the dihedral angle matches fairly well the atomistic reference, thus no optimization was performed. The angle distribution for the CG model has the same average as the atomistic target, but the width is reduced. This choice was required to ensure numerical stability with a time-step of 20 fs, as verified in runs with 370 PEO chains of length 20 over 900 ns (totaling over 7000 dihedral potentials, much larger than the system used in previous tests). The parameters chosen here represent a reasonable compromise between accuracy (with respect to reproducing atomistic distributions) and numerical stability.

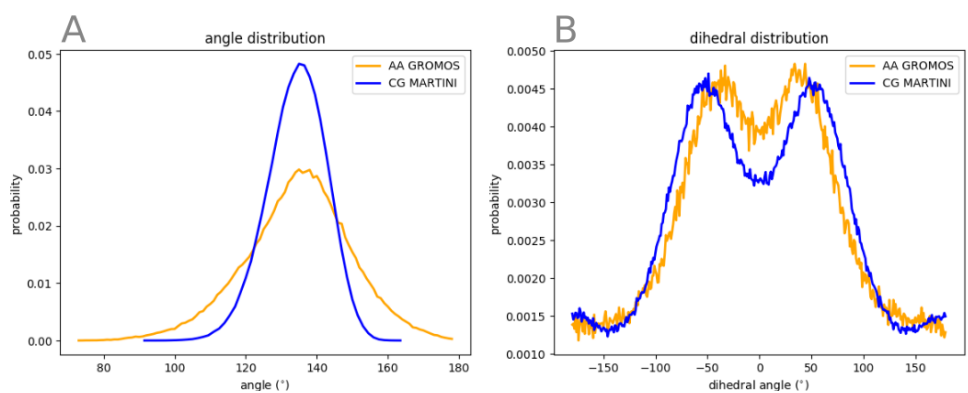

Figure 2. Comparisons of angle (A) and dihedral (B) distributions of PEO from atomistic and coarse-grained simulations.

## 4.2 - Validation of the new model

In order to demonstrate the transferability of the model and assess the range of molecular weights over which it can be applied, we performed a number of tests on PEO and PEO-containing compounds, considering five different application areas.

### 4.2.1 – PEO oligomer phase behavior and density

Phase behavior of PEO oligomers was not used as a target property during the parameterization of the new model. However, we checked that it is correctly reproduced for the specific case of triglyme (PEO tetramer). Simulations of water/triglyme mixtures (at 303.15 K and 1 bar pressure) were carried out at 6 different concentrations in the range from 10 mol% PEO to 80 mol% PEO. In this concentration range no demixing is observed, consistently with experiments.[41]

Table 4 shows the density of these mixtures measured in experiment, simulation using the Lee model and simulation using our model. Both MARTINI models deviate less than 5% from the experimentally measured values and our model improves over the Lee model in the high concentration regime. Moreover, we calculated the density of pure liquids for four short PEO oligomers, namely dimethoxyethane, diglyme, triglyme, and tetraglyme, at 298.15 K. The calculated densities agree fairly well with experimental values (Table 4), with a maximum deviation of 3%, lower than observed with the Lee model. Such agreement suggests that small PEO oligomers could be used as bulk solvents in MARTINI.

**Table 4. Densities for mixtures of triglyme (TIG) and water at 303.15 K as well as pure solutions of dimethoxyethane (DXE), diglyme (DGL) and tetraglyme (TRG) at 298.15.**

| Compound | mol% PEO | exp. | Lee model | present work** |
|---|---|---|---|---|
| DXE | 100 | 868.0* | 937.0* | 878.7 ± 0.4 |
| DGL | 100 | 945.0* | 1002.0* | 972.8 ± 0.3 |
| TRG | 100 | 1040.0* | 1067.0* | 1051.4 ± 0.3 |
| TIG | 100 | 975.86* | 1040.0 ± 0.3 | 1017.5 ± 0.3 |
| TIG | 80 | 979.80† | 1033.2 ± 0.3 | 1010.3 ± 0.4 |
| TIG | 50 | 990.01† | 1017.3 ± 0.3 | 994.4 ± 0.4 |
| TIG | 30 | 1002.98† | 1000.5 ± 0.3 | 978.1 ± 0.4 |
| TIG | 20 | 1012.61† | 988.1 ± 0.3 | 967.2 ± 0.3 |
| TIG | 10 | 1020.68† | 973.5 ± 0.3 | 957.9 ± 0.3 |

* taken from ref. 12

† taken from ref. 41

** unscaled densities are reported, because the real mass of the PEO monomer is about the same as the mass of the MARTINI EO bead.

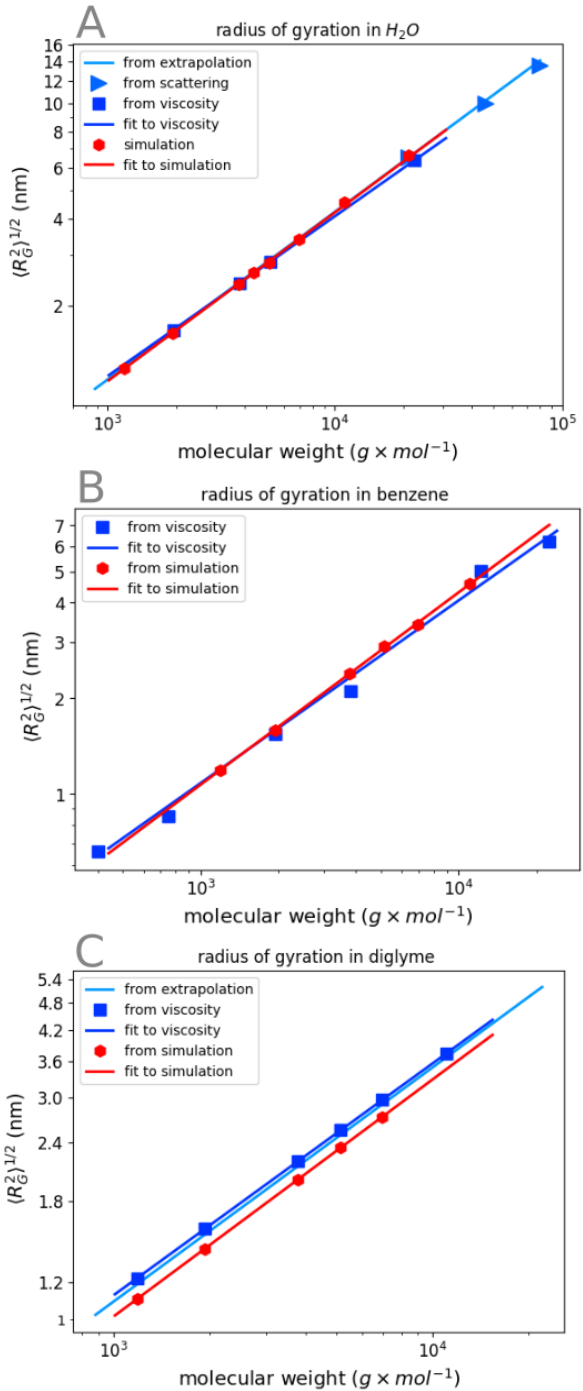

Figure 3. Radius of gyration for PEO as a function of molecular weight observed in different solvents: (A) water, (B) benzene, and (C) diglyme. Blue markers are experimental reference values obtained by the methods outlined in Supporting Information 3. The lines are extrapolations or fits to the corresponding (same color) data points.

## 4.2 - Long range structural properties

To obtain long range structural properties we simulated single PEO chains with different molecular weights, ranging from 1.2 kg/mol (~27 monomers) to 21 kg/mol (~477 monomers), in three different solvents. The simulations in water and benzene, which are both good solvents[39], were carried out at 298.15 K and 1 bar pressure. The simulation in diglyme (DGL) was run at 323.15 K (50°C), at which the otherwise bad solvent becomes a theta-solvent.[27]

### 4.2.3 - Radius of gyration

It is possible to compute the radius of gyration ($R_G$) directly from simulation data. $R_G$ is defined as the root mean square of the distance of all ($N$) atoms of the polymer chain from their center of mass ($CoM$).

$$<R_G^2>^{1/2} = \frac{1}{N}\sum_{k=1}^{N}(R_k - R_{CoM})^2$$

MARTINI polymer models are often parametrized not only to reproduce small oligomer free energies of transfer but also long range structural properties such as the radius of gyration ($R_G$).[2,4,8,18] $R_G$ for PEO-477 (20988 kg/mol) in water was our target during the parameterization stage. On the other hand, we also validate our model by comparing the $R_G$ for 6 other molecular weights in three different solvents to radii of gyration derived from experiment.

Comparing $R_G$ from experiment to simulation is not always straightforward. Experimental radii of gyration result from either direct or indirect measurements. Direct measurements, such as those obtained from light scattering, usually pertain to large molecular weights ($M_w > 100$ kg/mol), generally beyond those used in simulations. Hence, direct measurements can only be compared to simulation results by extrapolation. Indirect measurements, on the other hand, yield physical properties of polymer solutions at low concentrations. These properties, such as the intrinsic viscosity, can then be related to $R_G$ using theoretical or empirical models of real polymers. Intrinsic viscosity measurements are accurate and possible also for lower $M_w$ (> 1 kg/mol), comparable to the PEO chains simulated here. To validate our model, we used both approaches: extrapolation of $R_G$ from direct measurements, and estimation of $R_G$ from intrinsic viscosity data. The details of both approaches are reported in the Supporting Information (S.3).

All radii of gyration obtained by simulation with our new PEO model in comparison to experimental reference data are shown in figure 3. For water (panel A) the experimental reference data consists of an extrapolation from high molecular weight scattering data[42] (dashed blue line figure 3), three single points (blue triangles figure 3) measured by scattering experiments at high molecular weight[40], and estimates based on low molecular weight intrinsic viscosity measurements.[42] All three data sets agree well with each other and also with the radius of gyration produced by our model. For the three molecular weights (2.0 kg/mol, 3.784 kg/mol, 5.148 kg/mol) for which both simulation data and an estimate from viscosity data exists, a direct comparison can be made. For the two high molecular weights, the radius of gyration from simulation matches the one estimated from experiment within the standard error. At the lowest molecular weight, the match is not exact but the deviation is only 3% (see table S.3.1).

Using Flory theory to estimate the free energy of a polymer chain in a good solvent, it can be shown that the scaling relationship between radius of gyration and molecular weight is a power law, i.e., $R_G \propto M_W^\alpha$, with α=0.6. However, a more sophisticated theoretical treatment gives an exponent of 0.588.[28] Fitting experimental scattering data[42] to the same power law yields an exponent of 0.58, and fitting the data from the new MARTINI model yields 0.583 ± 0.002; this value is in excellent agreement with both the theories and the experimental data at higher molecular weight. On the other hand, we notice that a power law fit to the radii of gyration estimated from intrinsic viscosities yields an exponent of 0.552 ± 0.002. Considering that our model reproduces well the values of $R_G$ from viscosity estimates, extrapolation and scattering over almost two orders of magnitude of $M_w$, we consider this deviation in scaling negligible. Overall the agreement of our model and the experimental reference data is very satisfactory.

Figure 3B shows the radii of gyration for PEO in benzene based on estimates from intrinsic viscosity measurements, and based on simulations with our new model. In general the simulated values (red hexagons) are close to the reference values (blue squares). A direct comparison between the two is only possible for two molecular weights: 3.784 kg/mol (A) and 2.0 kg/mol (B). At the high molecular weight (A) we obtain a radius of gyration of 2.40 ± 0.02 nm from simulations and a radius of gyration of 2.11 ± 0.01 nm from experiment. The difference between the two values is 0.26 nm, corresponding to a relative deviation of about 14%. For the shorter PEO chain (B), the CG model yields a radius of gyration of 1.585 ± 0.005 nm, which is much closer to the experimental value (1.539 ± 0.007 nm); the relative deviation is about 3%. Overall, the deviation of our new model from the experimental reference values appears to be acceptable, bearing in mind that the error for the reference radii of gyration is only a lower bound to the real error (see also S.3). In addition, the CG model produces a scaling exponent of 0.603 ± 0.004, in perfect agreement with the expected value for a good solvent and with the fit of the experimental data (0.602 ± 0.003).

Figure 3C shows the radii of gyration for PEO in diglyme based on estimates from intrinsic viscosity measurements, based on extrapolation, and obtained from simulations with our new model. The estimates from intrinsic viscosity data (blue squares) are very close to the extrapolation from scattering data (blue dashed line). In contrast, at a molecular weight of 3.784 kg/mol, the CG model predicts a radius of gyration of 2.002 ± 0.007, whereas the estimated reference value is 2.189 ± 0.009. The deviation is about 10%, similar to the deviation obtained at other molecular weights (see table S.3.1). Such deviation can be ascribed in part to the higher temperature, at which the performance of our CG model becomes worse. However, overall the results from our model are in reasonable agreement with both experimental data sets over the entire range of molecular weights. In addition, the scaling exponent for the CG model is 0.511 ± 0.003, in good agreement with experimental data (0.505 from scattering measurments[42]) and close to 0.5, which is the value predicted by ideal chain statistics for a chain in a theta solvent.[28]

### 4.2.2 End-to-End distance and Kuhn Length

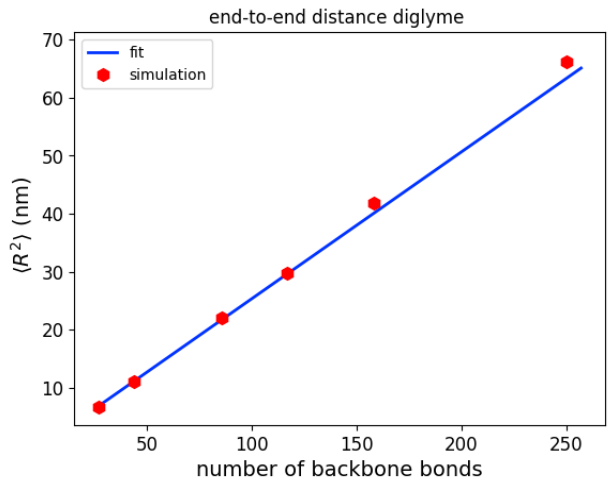

Figure 4. End-to-End distance (red hexagons) and fit (blue line) for the new MARTINI model as function of the number of backbone-bonds (i.e. 1 in case of our CG model) in diglyme, which is a theta solvent.

It can be shown from statistical calculations that, for sufficiently long chains, the squared end-to-end distance of a polymer chain ($<R^2>$) in a theta solvent is proportional to the bond length ($l$), the number of back-bone bonds ($n$), and a constant referred to as Flory's Characteristic ratio ($C_\infty$):[28]

$$<R^2> = C_\infty \times n \times l^2$$

Diglyme at 50 °C is a theta solvent for PEO[27] – in good agreement with our simulations, which is indicated by the scaling exponent of about 0.5. Therefore, $C_\infty$ can be obtained by fitting the above relation to the squared end-to-end distance of our model in diglyme (see Figure 4). From this procedure we yield a value for $C_\infty$ of 2.45 ± 0.002 nm, close to the value previously reported for the Lee model[6] (2.7 nm). The value of $C_\infty$ is related to the Kuhn length ($b$) and to the persistence length ($l_p$), which both can be interpreted as a measure of the polymer stiffness. From $C_\infty$ we can compute the persistence length according to[6]:

$$l_p = C_\infty \times \frac{1}{2} \times l$$

or the Kuhn length[28] following the equation:

$$b = \frac{C_\infty \times l}{\cos\theta}$$

The bond length ($l$) is given by 3.22 Å and the angle $\theta$ in our model is 45 degrees. This leads to a persistence length of 3.95Å, which compares fairly well to the experimental value of 3.7 Å.[6] Similarly, the Kuhn length of 11.56 Å calculated from simulations compares well with the value measured in the PEO melt (11.0 Å[28]).

### 4.2.5 - PEGylated lipids

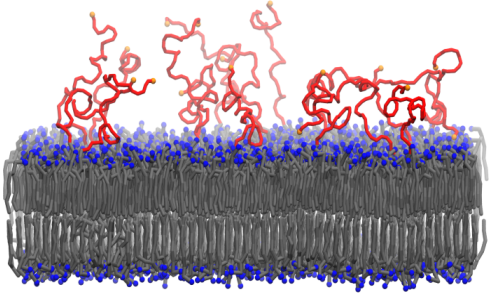

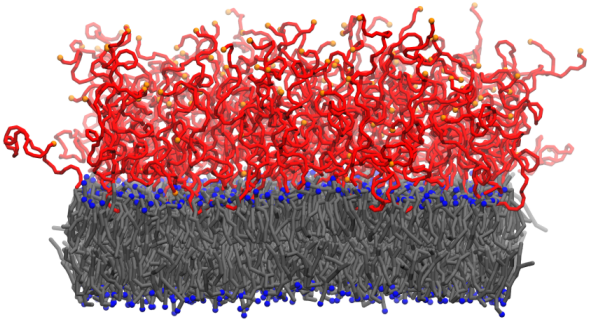

Figure 5. Lipid bilayers at low (A) and high (B) concentration of PEGylated lipids after 4 μs. Acyl chains are colored in gray, choline head groups in blue, PEO chains in red and terminal SP2 bead in orange. Water is not represented for the sake of clarity.

PEGylated lipids are interesting from the pharmaceutical standpoint for their applications in drug delivery.[43] From a polymer physics standpoint, membranes containing PEGylated lipids mimic PEO grafting to a solid surface. Chain dimensions in grafted polymers can be sorted into two regimes depending on the grafting density. In the low grafting density regime, known as "mushroom regime", the chains are well separated, interact minimally, and therefore move freely within a space approximating a half-sphere.[43] In contrast in the high grafting density regime ("brush regime"), the polymer chains are close in space and repel each other. This repulsion leads to more extended chain dimensions than in the mushroom regime. To verify that our model can reproduce this difference in dimension and therefore can be used as a new model for PEGylated lipids, we simulated two patches of PEGylated lipid-bilayers in water.

To obtain such patches, first a single chain of PEO was grown on top of a single lipid in a small lipid-bilayer using the Polyply package. Subsequently this small patch was replicated in x and y dimensions and equilibrated at high temperature (350K) for a short time. It was then cooled down to 283K to reach the same conditions as experimentally used for liposomes containing PEGylated lipids.[43] The grafting densities were 0.034 nm$^{-2}$ and 0.42 nm$^{-2}$; the areas per lipid were 0.46 nm$^2$ and 0.598 nm$^2$ respectively. Although the first bilayer was in a gel state, the second bilayer appeared to be in the liquid crystalline state. We do not expect the phase of the bilayer to have an effect on the PEO chain dimensions.

By visually inspecting the PEO chains, the dimensions look like the anticipated mushroom at low grafting density (panel A figure 5) and like the anticipated brush at high grafting density (panel B figure 5). To quantify the difference in dimension the end-to-end distance was computed. In case of the low grafting density, a value of 4.44 ± 0.02 nm is obtained, significantly smaller than the end-to-end distance in the brush regime (5.21 ± 0.02 nm). The chains in the brush regime are more extended than in the mushroom regime, as expected from theory and observed experimentally.[43–45]

As evident from chain dimensions, our new model for PEGylated lipids does not show strong adsorption of the PEO chains onto the lipid bilayer surface, contrary to the first Lee model.[7] Although some other coarse-grained models show enhanced adsorption onto the lipid bilayers, data from both atomistic simulations and experiment suggests that the PEO chains should not do so.[43] Thus our model displays the correct qualitative behavior without refinement.

The chain dimensions in the mushroom regime can be assessed in relation to experiment by estimating the end-to-end distance from an experimentally accessible parameter: the Kuhn length. According to Flory theory, the end-to-end distance for an isolated chain grafted to a surface (i.e., mushroom regime) is given by:[28]

$$R_F = b^{\frac{2}{5}} \times \left(3 \times n_{monomers} \times l \times cos\frac{\theta}{2}\right)^{\frac{3}{5}}$$

with $b$ being the Kuhn length, $n$ the number of bonds per chain, and $l$ the average bond length. The angle $\theta$ is 180 degrees minus the average angle between two consecutive bonds. Details on deriving the Kuhn length from experiment and the other quantities are presented in the supporting information (S.4). For the low grafting density (σ=0.034), the end-to-end distance from our simulation (4.44 ± 0.02 nm) compares very well to the estimated value of $R_F$ (4.8 ± 0.4) nm.

Similarly, the chain dimensions in the brush regime can be assessed by estimating the height of the brush ($H$) in terms of the Kuhn length and the grafting density (σ). Using Alexander - de Gennes theory, the height is given by:[28]

$$H = 3 \times n_{monomers} \times l \times cos\frac{\theta}{2} \times b^{\frac{2}{3}} \times \sigma^{0.65}$$

where σ is the number of grafting points per unit area (i.e. the number of PEGylated lipid per unit area). Supporting information S4 offers more details on the approximations and quantities involved in this equation. The estimated brush height is 8.0 ± 2 nm.

To define the height of the brush from simulation is somewhat more difficult, as there is no well accepted procedure. Previously, Lee *et al.* have used the peak of the density profile computed with respect to the choline head group as the height of the brush.[7] Such profile is shown in figure 5 (red line). The peak is located at 3.3 nm, smaller than the end-to-end distance and much smaller than the estimate based on Alexander-de Gennes theory (8 ± 2 nm). In contrast, the peak (7.5 ±1.0 nm) of the density profile of only the SP2 chain-end beads (red line in figure 5), reminiscent of the less dense brush top, is in good agreement with the estimated value. Yet a third measure of the brush height in simulation could be the average end-to-end distance (5.21 ± 0.02 nm), which lies in-between the two previous measures and is reasonably close to the estimated value. Overall, a direct comparison between simulation results and Alexander-de Gennes theory appears to be problematic, possibly due to the assumption (in the theory) of idealized straight chains.

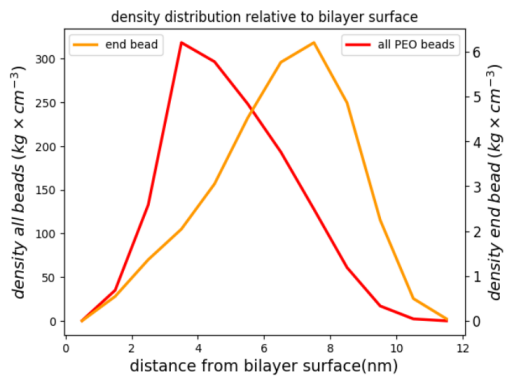

Figure 5. Density profile of the PEO beads (red) and only the SP2 beads (orange) with respect to the bilayer surface, taken as the choline head group (indicated in blue in figure 4).

In conclusion, the new model (1) can reproduce the difference in size of PEO chains in PEGylated lipids, both at low and high grafting densities; (2) does not suffer from artificially high aggregation of chains at the bilayer surface; (3) yields reasonable chain sizes in comparison to experimental estimates.

### 4.2.6 - Nonionic surfactants

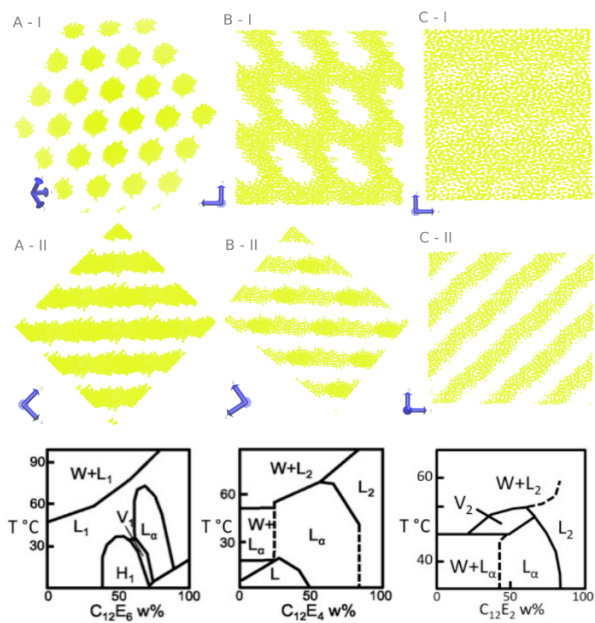

Figure 6. Snapshots of non-ionic surfactant phases after 5μs microseconds and corresponding phase diagrams. Only the carbon atoms are shown. Panels A correspond to C12E6, panels B to C12E4 and panels C to C12E2. Panel A-I shows the phase looking down the edge of a cube, A-II is rotated looking along the edge of the cube. Panel B-I shows the holes in the lamellae looking down the face, while panel B-II shows that the lamellae are indeed separated. Panel C-I shows no holes in the lamellae and panel C-II the corresponding separated sheets.

Non-ionic surfactants with a PEO head group and an alkyl tail are another important application of PEO. In water, they display a rich phase behavior. In simulations, such phase behavior is very sensitive to both bonded and non-bonded interactions. Hence, the phase behavior of non-ionic surfactants is an ideal, stringent test for any PEO model. Because of the richness of phase behavior in non-ionic surfactants, we selected only 3 specific cases, in which surfactants produce different morphologies; these specific cases are the same selected in the previous work by Rossi *et al.*, to simplify the comparison. [8]

Three non-ionic surfactants, namely C12E6, C12E4 and C12E2, were simulated in water at three different concentrations, corresponding to unambiguous regions in their respective phase diagrams (Figure 5, lower panels). Surfactant molecules were initially distributed randomly in the simulation box. Self-assembly simulations were repeated 6 times for each surfactant and each concentration, each time with a different random-seed to generate different random velocities from a Maxwell distribution at the appropriate temperature.

C12E6 was simulated at 50% (w/w) water content. At the temperature of 298.15 K, the phase diagram indicates that a hexagonal phase should form. In self-assembly simulations, we obtained tubular micelles in 5 out of 6 cases, and in 3 out of 6 the tubes have hexagonal symmetry (Figure 5, panel A). Only 1 out of 6 simulations yielded an unidentifiable phase.

C12E4 at 53% (w/w) water at the temperature of 298.15 K forms lamellar phases, according to the experimental phase diagram. In self-assembly simulations (Figure 5, panel B-I and B-II), we obtained lamellar structures in 4 out of 6 cases (the remaining 2 simulations gave tubular micelles). However, the lamallae showed holes. Whereas the holes could not be identified in x-ray scattering experiments, more recent NMR data clearly shows that the order parameter is not compatible with intact lamellar phases, and instead are compatible with a perforated lamellar phase.[46] We notice that the Rossi model also produced perforated lamellae (3 cases out of 6). [8]

Finally, C12E2 at 71.1% (w/w) water content and T=298.15 K forms lamellar phases, according to the phase diagram. In self-assembly simulations, the surfactant formed intact lamellae in 6 out of 6 cases (Figure 5, panel C-I and C-II).

Overall, the new model is able to predict the correct phase behavior as observed experimentally in the three cases tested here. Also, in such cases, the agreement with experimental phase behavior is better than observed for the previous PEO model.[8]

### 4.2.7 - PS-PEO block-copolymer aggregates

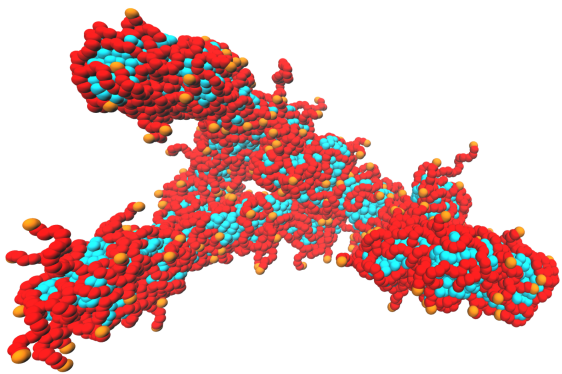

Figure 6. PS-PEO block copolymer micelle (219 oligomers) in water after 3.4 microseconds. The PEO part is shown in red while the PS part is cyan and water is omitted for clarity.

PEO is not only frequently used as component of biomolecular systems, but is also a very important polymer in material science and engineering. Recently the block-copolymer of polystyrene (PS) with PEO (PS-b-PEO) has attracted some attention in theoretical studies on Lithium ion conducting polymers.[47–50] Short oligomers of PS-b-PEO form micelles in water. The size and the aggregation number of these micelles have been characterized by x-ray scattering.[51] We tested the possibility to combine the new PEO model with the existing MARTINI model of PS,[2] by simulating two systems of 370 and 740 PEO-b-PS oligomers in water; each chain contained 23 consecutive PEO units and 10 consecutive PS units.

In the beginning of both simulations, small micelles were formed, which later fused to generate bigger micelles. The PEO part of the polymer wrapped around the PS, creating a PS core and a PEO corona, as to shield PS from water. For the smaller system, after 2.5 μs only 4 micelles remained (formed by 15, 34, 102 and 219 oligomers, see figure 6). For the larger system, after the same period of time, 9 micelles remained (formed by 456, 49, 37, 36, 64, 15, 41, 23, 20 oligomers). These micelles were stable for about 1 μs in both cases. The aggregation number of the largest micelle does not match the experimentally determined aggregation number (370 oligomers[51]), in both of our simulated systems. At the same time, the radius of gyration of the largest micelle for the first system (6.993 nm) and for the second second system (8.991 nm) fall in the same ballpark as the experimentally determined value (6.3 ± 1 nm[46]). The discrepancy in micelle size may simply be due to time scale limitations: fusion of smaller micelles with the larger ones or division of larger micelles into smaller ones probably occurs on time scales larger than those accessible in our simulation. In addition, the large gap between the aggregation number of the small micelles and the large one suggests that the larger one is favored and kinetic barriers prevent further fusion or division. Length scale limitations may also play a role: in real systems, micelles are polydisperse, i.e., they have a range of different sizes and aggregation numbers, and exchange monomers dynamically; in simulations, such dynamic equilibrium would imply system sizes currently out of reach, even for coarse-grained models. We note that previous studies of micelle formation (with other surfactants[7,10]) using MARTINI models also yielded only qualitative agreement with experiment. Our results indicate that the new PEO model can be combined with the existing PS-model without modifications.

### 5.0 Conclusions

Motivated by the deficiencies of the previous MARTINI PEO models in apolar environments, we developed a new PEO model based on (a) a set of 8 free energies of transfer of dimethoxyethane (PEO dimer) from water to solvents of varying polarity; (b) the radius of gyration of a PEO-477 chain in water at high dilution; (c) matching angle and dihedral distributions from atomistic simulations. The radius of gyration for PEO chains of different length in different solvents was not used in the parameterization, but it turned out to be in good agreement with experiments. We showed that the model can be used on a molecular weight range from about 1.2 kg/mol to 21 kg/mol (27 to 477 monomers) and possibly even higher. The new model successfully reproduces the phase behavior and densities of small PEO oligomers in water. It can be used in polar as well as apolar solvents, such as benzene. We also verified that the new model can be used as part of PEGylated lipids, and reproduces qualitatively the structural features of the lipid bilayers with PEGylated lipids in the brush and mushroom regime. Furthermore, the model is able to reproduce the phase behavior of various non-ionic surfactants in water, even improving on the Rossi model. Finally, we demonstrated that the new model can be combined with the existing MARTINI PS to model PS-PEO block-copolymers. In conclusion, the new parameterization captures all the essential properties of the previous models and improves on their deficiencies, yielding a highly transferable and stable coarse grained model for PEO.

**Supporting Information**

Topologies with details on bonded interactions of the PEO derived compounds (PEGylated lipids, non-ionic surfactants and PS-b-PEO copolymers) are provided online (http://mmsb.cnrs.fr/en/team/mobi). All scripts and programs used are also available online (https://github.com/fgrunewald/Martini_PolyPly and https://github.com/fgrunewald/tools_for_MD_analysis).

### ACKNOWLEDGMENT

LM acknowledges the Institut national de la santé et de la recherche médicale (INSERM) for funding. Calculations were performed at the French supercomputing center CINES (grant A0020710138) and on the Peregrine high performance computing cluster of the University of Groningen. We would like to thank the Center for Information Technology of the University of Groningen for their support and for providing access to the Peregrine high performance computing cluster.

FG acknowledges financial support by the EACEA in form of an Ersasmus+ scholarship (project no.: 159680-EPP-1-2015-ES-EPPKA1-EPQR). GR acknowledges funding from the ERC Starting Grant BioMNP – 677513.